\documentclass[aps,prb,showpacs,twocolumn]{revtex4}

\usepackage{graphicx}

\begin{document}
\draft

\title{Non-damping magnetization oscillations in a single-domain ferromagnet}

\author{P.~M.~Gorley$^1$, P.~P.~Horley$^1$, V.~K.~Dugaev$^{2,3}$, J. Barna\'s$^4$,
and V. R. Vieira$^2$}
\address{
$^1$Department of Electronics and Energy Engineering, Chernivtsi National
University, 2 Kotsyubynsky St., 58012 Chernivtsi, Ukraine\\
$^2$CFIF and Department of Physics, Instituto Superior T\'ecnico,
Av. Rovisco Pais, 1049-001 Lisbon, Portugal,\\
$^3$Frantsevich Institute for Problems of Materials Science,
National Academy of Sciences of Ukraine, 5 Vilde St., 58001 Chernivtsi,
Ukraine\\
$^4$Department of Physics, Adam Mickiewicz University, Umultowska~85, 61-614~Pozna\'n, and\\
Institute of Molecular Physics, Polish Academy of
Sciences, M.~Smoluchowskiego~17, 60-179~Pozna\'n, Poland}
\date{\today }

\begin{abstract}
Non-damped oscillations of the magnetization vector of a
ferromagnetic system subject to a spin polarized current and
an external magnetic field are studied theoretically by solving the
Landau-Lifshitz-Gilbert equation. It is shown that the frequency
and amplitude of such oscillations can be controlled by means of
an applied magnetic field and a spin current. The possibility of
injection of the oscillating spin current into a non-magnetic
system is also discussed.
\end{abstract}
\pacs{85.75.-d, 72.25.-b, 05.65.+b, 05.45.-a}

\maketitle

Spin polarized current incident on a magnetic system can exert a
torque on its magnetic moment. This torque, in turn, can change a
magnetic state of the system. One possibility is a switching from
a certain magnetic configuration to another one, as predicted
theoretically~\cite{slonczewski96} and also observed
experimentally in several spin valve
structures.\cite{katine00,grollier01,darwish04} The phenomenon of
current-induced magnetic switching is a consequence of the spin
transfer from the conduction electron system to the localized spin
moments.\cite{slonczewski96,stiles02} In certain circumstances,
however, the spin transfer torque can induce some stable
non-damped precessional modes. In such states, the energy is
pumped from the spin current to the localized moments, which support
the magnetization precession. Such non-damped precessions are of high
importance from the point of view of possible applications in
the microwave generation.\cite{kiselev03,krivorotov05,xiao05}

Another important issue in spintronics is the spin injection from
ferromagnetic to nonmagnetic metals (and/or semiconductor) and
the spin control over distances comparable to the spin diffusion
length.\cite{dietl02,ferrand01,rashba00} Materials which might be
promising for the applications in spintronics should have a relatively
long spin diffusion length (of the order of the system size) and
also should allow an efficient spin injection across interfaces.
Despite several technological and fundamental problems, there is
some progress concerning the efficiency of the spin injection and its
control by some external parameters.

From the physical point of view, the phenomena of the spin
transfer torque and the spin injection are not independent. This
is because the current-induced switching relies on the spin
coherence between two magnetic bodies across a nonmagnetic spacing
material. In this paper we consider precessional modes of a
ferromagnetic system, driven by a spin polarized current, and the
associated injection of circularly polarized electrons into a
nonmagnetic system. This is an extension of our earlier work,
where we have studied equilibrium and stationary states of such a
system.\cite{gorley05} To study the magnetic dynamics of the
system we used the Landau-Lifshitz-Gilbert equation, with the spin
transfer torque included. We also assumed that the torque is an
interfacial effect, i.e., the component of the spin current
perpendicular to magnetization is absorbed at the very
interface.\cite{sun00,brataas01,barnas05} Here we use the same
model and description to study stable precessional modes.

The time variation of the angular variables $\theta$ and
$\varphi$, which characterize an orientation of the unit vector $\bf
m$ along the magnetization of the ferromagnetic system, can be
written in dimensionless variables as~\cite{gorley05,sun00}
\begin{eqnarray}
\label{Eq1}
\frac{\partial \theta}{\partial \tau} = -\sin \theta \left[ \alpha
\left(Z(\varphi)\cos \theta + h\right) + \frac{h_p}2 \sin 2\varphi + h_s\right], \\
\frac{\partial \varphi}{\partial \tau} = - \left[ Z(\varphi) \cos \theta + h -
\alpha \left( \frac{h_p}2 \sin 2\varphi + h_s \right) \right], \nonumber
\end{eqnarray}
with the dimensionless time defined as
$\tau=t/[(1+\alpha^2)/\gamma H_k )]$, where $\gamma$ is the
gyromagnetic ratio, $H_k$ is the anisotropy field, and $\alpha $
is the damping coefficient. Apart from this, $h$, $h_s$, and $h_p$
are the dimensionless external magnetic field, spin current, and
the easy-plane anisotropy field, respectively, defined as in
Ref.~[\onlinecite{gorley05}]. Finally, $Z(\varphi) \equiv 1+h_p
\cos ^2 \varphi$. Equations (\ref{Eq1}) are written for the case
when the magnetic field and spin current are collinear with the easy
axis of the ferromagnetic system.

In the stationary case ($\tau \to \infty$), the system (\ref{Eq1})
can be transformed into a set of two trigonometric equations with
respect to the angles $\theta_0$ and $\varphi_0$, with the general
periodic solutions:
\begin{eqnarray}
\label{Eq2}
\sin 2\varphi_0 = -2h_s/h_p, \hskip0.5cm
\cos\theta_0 = -h / Z(\varphi_0).
\end{eqnarray}
The invariance properties for Eqs.~(\ref{Eq1}) allow to select only two
independent stationary states of Eqs.~(\ref{Eq2}), which can be
written in the form~\cite{gorley05}
\begin{eqnarray}
\label{Eq3}
\varphi_{01}+\varphi_{02}=\pi/2,\hskip1cm
\nonumber \\
\cos \varphi_{02} = - {\rm sgn}\, h_s
\left[ \frac12 \left( 1-\sqrt{1-4h_s^2 /h_p^2}\right) \right] ^{1/2},\\
\sin \varphi_{02} = \left[ \frac12 \left( 1+\sqrt{1-4h_s^2 /
h_p^2}\right) \right] ^{1/2}, \nonumber \\
\sin \varphi_{01}=\cos
\varphi_{02}, \hskip0.5cm \cos\varphi_{01}=\sin\varphi_{02},\nonumber \\
\cos\theta_{01,2}=-h/ Z(\varphi_{01,2}). \nonumber
\end{eqnarray}

As follows from Eqs.~(\ref{Eq3}), both stationary states and their
energies do not depend on the damping coefficient $\alpha$, and
are determined by $h$, $h_p$ and $h_s$ only. Moreover, the latter
parameters satisfy the condition $|h_s| \le 0.5 h_p$.

To investigate the stability of the stationary solutions (\ref{Eq3}),
we subject them to a small time-dependent perturbation
\begin{eqnarray}
\label{Eq4}
\theta~=~\theta_0 + \delta \theta \; e^{-i\omega \tau},
\hskip0.2cm \varphi~=~\varphi_0+\delta\varphi\; e^{-i\omega \tau},
\\
\delta\theta = {\rm const} \ll \theta_0, \hskip1cm \delta\varphi
= {\rm const} \ll \varphi_0, \nonumber
\end{eqnarray}
with $\omega$ being generally a complex variable,
\begin{equation}
\label{Eq5}
\omega~=~\omega_r + i\, \Delta\omega,
\end{equation}
where $\omega_r$ is the frequency of homogeneous precession of the
magnetic moment in an intrinsic effective magnetic field, whereas
$\Delta\omega$ describes damping (or growth) of the soft mode
fluctuation amplitude and corresponds to the natural width of
ferromagnetic resonance band.

Substituting Eq.~(\ref{Eq4}) into Eqs.~(\ref{Eq1}) and applying the
standard linearization procedure~\cite{erugin74} with respect to
the perturbation, one obtains the characteristic equation of the
second order in $\omega$, which -- on taking into account
Eq.~(\ref{Eq5}) -- yields the following solutions depending on the
sign of $\Delta = (a_{11}-a_{22})^2 + 4a_{12}a_{21}$:
\begin{eqnarray}
\label{Eq6} {\rm a)}\;\;~\Delta>0:
~\omega_r=0,~\Delta\omega=0.5\left( a_{11}+a_{22}\pm\sqrt{\Delta}\right) , \\
\label{Eq7} {\rm b)}\;~\Delta<0:~\omega_r=\pm 0.5
\sqrt{-\Delta},~\Delta\omega=0.5\left( a_{11}+a_{22}\right) .
\end{eqnarray}
The matrix elements $a_{ij}$ in Eqs.~(\ref{Eq6}) and (\ref{Eq7})
are defined by the stationary solutions (\ref{Eq3}) as
\begin{eqnarray}
\label{Eq8}
a_{11}=\alpha Z(\varphi_0)\, \sin^2\theta_0, \hskip0.5cm
a_{21}=Z(\varphi_0)\, \sin \theta_0,\nonumber \\ a_{12}=h_p \sin \theta_0
\left( \alpha\cos \theta_0 \sin 2 \varphi_0 - \cos 2 \varphi_0\right) ,\\
a_{22}=h_p\left( \cos\theta_0 \sin 2 \varphi_0 + \alpha \cos 2 \varphi_0\right) .\nonumber
\end{eqnarray}

As follows from Eqs. (\ref{Eq6}) and (\ref{Eq7}), the oscillatory
states in the system are possible only for $\Delta < 0$. The
propagation of the non-damped oscillations of the magnetization
components is possible when the conditions $\Delta < 0$ and
$a_{11}=-a_{22}$ ($\Delta\omega=0$ in (\ref{Eq7})) are
simultaneously obeyed, which takes place for
$h_{s}(\Delta\omega=0) \equiv h_{sb}$, corresponding to the
solution of the equation
\begin{equation}
\label{Eq9}
h_{sb}^4 - a_1 h_{sb}^3 + a_2 h_{sb}^2 + a_3 h_{sb} - a_4 =0,
\end{equation}
with the coefficients
\begin{eqnarray}
\label{Eq10}
a_1=\frac{4h}{3\alpha}, \hskip0.7cm
a_3 = \frac{4h}{9\alpha} \left( 1+ h_p +h_p^2 - h^2 \right), \nonumber
\\ a_2=\frac{4}{9} \left[ \frac{h^2}{\alpha^2} + 2.5(1+h_p)-0.5h_p^2 +1.5h^2
\right], \\a_4 = \frac{1}{9} \left( -1 + h_p +h^2\right) \left( 1 + 3h_p
+2h_p^2 - h^2\right). \nonumber
\end{eqnarray}

Equation (\ref{Eq9}) is invariant with respect to the simultaneous
change of the sign of $h$ and $h_{sb}$, $h \to -h$ and $h_{sb} \to
-h_{sb}$, which corresponds to one of the invariance properties of
Eq.~(\ref{Eq1}).\cite{gorley05} Our analysis has shown that
Eq.~(\ref{Eq9}) has one real physical solution.

The derived formulas (\ref{Eq7}) to (\ref{Eq10}) allow to calculate
a characteristic dependence of the non-damped oscillations of the
magnetization vector on the control parameters $\alpha$, $h_p$,
$h_s$ and $h$ (in the linear approximation regarding the
perturbation). In our calculations we assumed $\alpha$=0.005 and
$h_p$=5 (as in Refs.~[\onlinecite{sun00,gorley05}]), reducing in this
way the number of control parameters to $h$ and $h_s$.

Let us consider now the behavior of $| \omega_r |$ and $| \Delta\omega
|$ with $h$ and $h_s$  (Fig.~1). For a better presentation, the
values of $| \Delta \omega |$ were multiplied by a factor of 30
in Fig.~1. As one can see, $\omega_r$ changes slightly with $h_s$
and depends quadratically on $h$, reaching a maximum value at
$h=0$. The surface $\Delta \omega =f(h, h_s)$, in turn, depends in
a more complex way on both $h$ and $h_s$. It is important to note
that $\Delta \omega$ may take either positive or negative values,
which correspond to an increase or decrease  in time of the
perturbation amplitude. When $\Delta \omega =0$, the system is
turned to the neutral mode, when non-damped oscillations of a
constant amplitude propagate through the system. As follows from
our calculations, decrease of the planar anisotropy $h_p$ leads to
a further complication of the surfaces $\omega_r =f(h, h_s)$ and
$\Delta \omega =f(h, h_s)$, while increase of $h_p$ makes them
smoother. The increase of the damping coefficient $\alpha$ has
practically no influence on the $\omega_r =f(h, h_s)$ surface.

\begin{figure}\label{Fig1}
\includegraphics[scale=0.8]{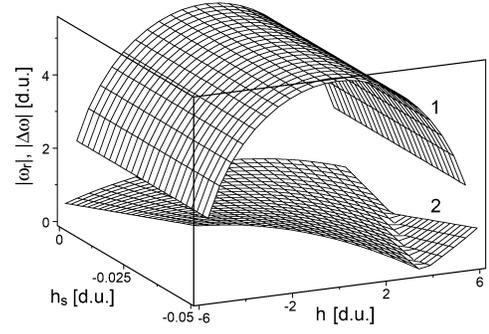}
\caption{ $|\omega_r|$ (surface 1) and 30$\times |\Delta \omega |$
(surface 2) as a function of $h$ and $h_s$  for the stationary
state ($\theta_{01}$, $\varphi_{01}$).}
\end{figure}

\begin{figure}\label{Fig2}
\includegraphics[scale=0.8]{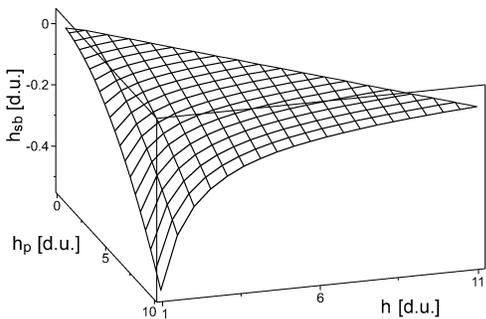}
\caption{ Spin current corresponding to the neutral, calculated as
a function of $h$ and $h_p$.}
\end{figure}

The non-damped oscillations in the system are possible only when
$1 < h < 1+h_p$ (see Ref.~[\onlinecite{gorley05}]). This condition
sets limits on the value of $h_{sb}$ as a solution of
Eq.~(\ref{Eq9}). Figure 2 presents a surface $h_{sb}=f(h, h_p)$
corresponding to the parameters at which the system is in a
state of non-damped magnetization oscillations, and thus being the
boundary between the spin-current stable ($h_s > h_{sb}$) and
unstable ($h_s < h_{sb}$) states. As follows from Fig.~2, $h_{sb}$
depends in a non-linear way on both arguments.

\begin{figure}\label{Fig3}
\includegraphics[scale=0.8]{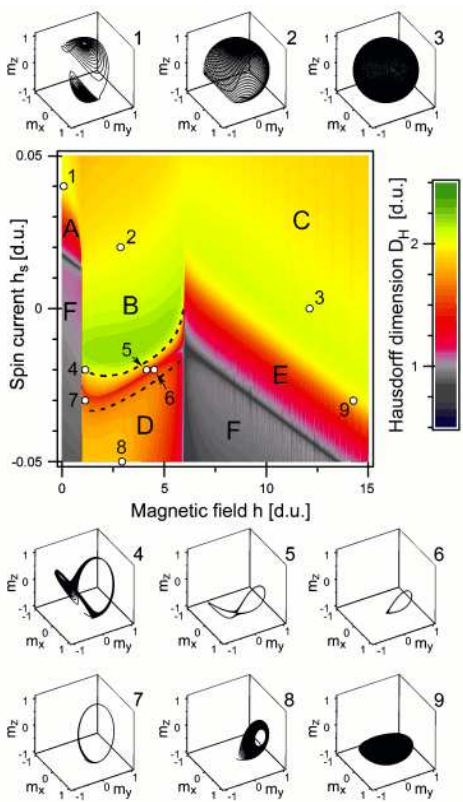}
\caption{(Color online) Hausdorff dimension diagram $D_H(h,h_s)$ for the system and the main phase
portrait types for different magnetic fields $h$ and spin
currents $h_s$: 1) $h=0.4,\, h_s=0.04$; 2) $h=2.825,\, h_s=0.02$;
3) $h=12.1,\, h_s=0$; 4) $h=1.1,\, h_s=-0.02$; 5) $h=4.125,\,
h_s=-0.02$; 6) $h=4.5,\, h_s=-0.02$; 7) $h=1.075,\, h_s=-0.03$;
8) $h=2.925,\, h_s=-0.05$; 9) $h=14.25,\, h_s=-0.03$. Dynamic
modes $A-F$ are described in the text.}
\end{figure}

Consider now numerical solutions of the nonlinear equations (1).
Figure 3 presents Hausdorff dimension\cite{haken83} diagram $D_H(h,h_s)$ showing main dynamical
modes of the system (phase states) for the different applied
fields and spin currents. The phase portraits corresponding to the
most characteristic points of the parameter space are shown above
and below the bifurcation diagram and numbered from 1 to 9. The
areas $A$, $B$ and $C$ correspond to the dynamic modes, for which
the transition from the initial ground state $m_z=-1$ to the state
$m_z=+1$ takes place. In the area $A$ the precession of the
magnetization vector takes place mainly along the axis $x$, and
the phase trajectory does not cover all the unit sphere (phase
portrait 1). On the contrary, the phase portraits 2 and 3,
characteristic to the areas $B$ and $C$, illustrate the transition of
the phase point to the upper pole {\it via} spiral trajectory,
with different initial behavior of the phase point running along
the two-loop curve (area $B$) or moving from the lower pole along
the spiral (area $C$). Under a constant spin current and
increasing magnetic field, the amplitude of the two-loop curve
decreases and the phase portrait 2 (area $B$) turns into that of
the phase portrait 3 (area $C$).

When the spin current decreases, the precession of the
magnetization vector slows down and the phase point becomes unable
to reach $m_z$=+1, remaining in the vicinity of the two-loop curve
and becoming a limit cycle (points 4 and 5 in Fig.~3, below the
upper dashed line and the corresponding phase portraits) for the
magnetic fields $1<h<1+h_p$. Under a further decrease of the spin
current, the limit cycle of the system turns into a single-loop
curve (phase 6 and 7), whose form and amplitude depend on the
control parameters $h$ and $h_s$. With a further decrease of the
spin current (area $D$), the limit cycle becomes unstable and the
magnetization vector instead of periodic non-damped oscillations
relaxes to a certain state with negative $m_z$ and zero $m_x$
(phase portrait 8). Upon approaching $h_{lim}=1+h_p$, the
non-stable cycle shrinks down to $m_z=-1$. It is worth noting that
the boundary between the areas $B$ and $D$ is well-defined and
sharp, contrary to the gradual transition between the areas $A-C$.

When the spin current decreases, the phase portraits corresponding
to the area $C$ keep the same oscillation type, but the
magnetization precession becomes significantly slower and the
resulting spiral trajectory covers only a part of the unit sphere
(phase portrait 9, area $E$). For the spin currents corresponding
to the area $F$ (for $h<1$ and $h>1+h_p$), the phase point is
unable to leave the ground state $m_z=-1$. Thus, the magnetization
vector can perform non-damped oscillations, forming the phase
portraits of a closed cycle for the narrow band of the control
parameter values (between the dashed curves around the boundary
between $B$ and $D$ in Fig.~3). Some examples of the time
dependence of $m_z$, illustrating non-damped oscillations for the
phase portraits 5, 6 and 7 are presented in Fig.~4. The curves in
Fig.~4 have been plotted starting from a certain time $\tau$ to
eliminate the influence of the transition processes taking place
in the vicinity of $\tau$=0. As can be seen in Fig.~4(a), for a
given $h_s$ and increasing $h$ one obtains oscillations of the
lower amplitude and frequency. Increase of $h_s$ (Fig.~4(b), solid
line) leads to the high-amplitude oscillation with smaller
frequency. This indicates the possibility of controlling the
period and amplitude of the non-damped oscillations of the
magnetization component $m_z$ by the magnetic field and spin
current.

\begin{figure}
\label{Fig4}
\includegraphics[scale=0.8]{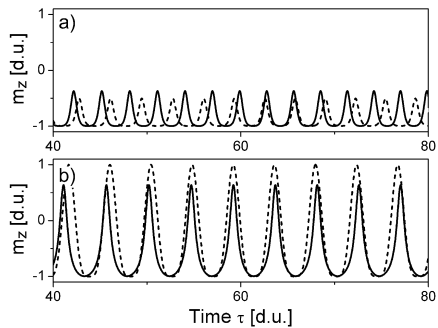}
\caption{ Time dependence of $m_z(\tau)$ for different magnetic
fields $h$ and spin currents $h_s$: a) $h_s=-0.02$, solid line
$h=4.125$, dashed line $h=4.5$; b) $h_s=-0.03$, solid line
$h=1.075$, dashed line - calculations according to (\ref{Eq13}).}
\end{figure}

To make a qualitative description of the $m_z(\tau)$ behavior
(Fig. 4) we use the expression for $\theta(\tau)$ from Ref.~[\onlinecite{gorley05}],
obtained in the linear
approximation with respect to the perturbation,
\begin{equation}
\label{Eq11}
\theta(\tau) = \pi + 2\pi \exp(\Delta \omega )\, \sin (\omega_r \tau).
\end{equation}
When writing Eq. (\ref{Eq11}), the initial condition $\theta(0) =
\pi$ was assumed. For the non-damped oscillation mode
($\Delta\omega = 0$), Eq.~(\ref{Eq11}) yields
\begin{equation}
\label{Eq12}
m_z(\tau)= -\cos \left[ 2 \pi \sin (\omega_r \tau )\right] .
\end{equation}

As one may expect, the latter equation, obtained in the linear
approximation, may not give good quantitative description of the
essentially nonlinear $m_z(\tau)$ behavior, presented in Fig.~4
as a result of numerical calculations. However, one can obtain
much better agreement by taking the function of the form
\begin{equation}
\label{Eq13}
m_z(\tau)= -\cos \left\{ 2 \pi \sin^2 \left[ \omega_{r0}(\tau + \tau_0)\right] \right\} ,
\end{equation}
where $\omega_{r0}$ and $\tau_0$ are some approximation
parameters. For example, the $m_z(\tau)$ curve calculated according to
(\ref{Eq13}) for $h_s=-0.03$,  $\omega_{r0}=1.553$ and
$\tau_0=-0.053$ (Fig.~4(b), dashed line) shows good agreement
with the corresponding curve obtained by numerical methods
(Fig.~4(b), solid line).

Equation (\ref{Eq13}) allows to perform a qualitative description (in the
first approximation) of the possible injection of
non-damped $m_z(\tau)$  oscillations from ferromagnetic ($F$) into
non-magnetic ($N$) material. In the case of an ideal injecting
contact at $x$=0, which does not change the value and orientation
of the spin, one may assume that the spin currents to the left
and to the right of the contact
are equal,\cite{rashba00} $h_s^F = h_s^N \equiv
h_s(0,\tau)$. In such a case, the functional dependence
$h_s(x,\tau)$ can be obtained from the continuity
equation~\cite{gregg02}
\begin{equation}
\label{Eq14}
\frac{\partial m_z}{\partial \tau} + \frac{\partial h_s
(x,\tau)}{\partial x}=0.
\end{equation}
Introducing (\ref{Eq13}) into (\ref{Eq14}), one can show that
\begin{equation}
\label{Eq15}
h_s(x, \tau) = h_s(0, \tau) + \pi \omega_{r0}(
\xi_+-\xi_-)\; x,
\end{equation}
where $\xi_\pm=\cos \left[ 2 \left\{ \pi \sin^2 \omega_{r0}(\tau + \tau_0) \pm
\omega_{r0}(\tau + \tau_0) \right\} \right] $.

As follows from Eq.~(\ref{Eq15}), the spin current injected from
the ferromagnetic into the nonmagnetic system will preserve its
non-damped oscillation character, being a superposition of the
second harmonics of two harmonic oscillations. In the framework
of the current assumptions it will change linearly in amplitude
with distance from the contact. It is worth noting that the
inclusion of relaxation item~\cite{gregg02} does not change
Eq.~(\ref{Eq15}) qualitatively. For resistive and other contact
types~\cite{rashba00, schmidt00,albrecht03}, the expression for
the spin current will differ from that given by Eq.~(\ref{Eq15}). The
results of this paper, however, show that the non-damped
oscillations of $m_z(\tau)$ component can be injected from the
ferromagnetic to a nonmagnetic system due to current continuity at
the contact.

\begin{acknowledgments}
This work is partly supported by FCT Grant POCI/FIS/58746/2004
(Portugal), EU RTN2-2001-00440 'Spintronics', and Centre of
Excellence. V.D. thanks the Calouste
Gulbenkian Foundation in Portugal for support .
\end{acknowledgments}


\begin{thebibliography}{99}

\bibitem{slonczewski96}
J. C. Slonczewski, J. Magn. Magn. Mater. {\bf 159}, L1 (1996);
{\bf 195}, L261 (1999); L. Berger, Phys. Rev B {\bf 54}, 9353 (1996).

\bibitem{katine00}
J. A.~Katine, {\it et al.,}
%F. J.~Albert, R. A.~Buhrman, E. B.~Myers, and D. C.~Ralph,
Phys. Rev. Lett. {\bf 84}, 3149 (2000).

\bibitem{grollier01}
J. Grollier, {\it et al.,}
%V. Cros, A. Hamzic, J.M. George, H. Jaffres, A. Fert, G. Faini,
%J. Ben Youssef, and H. Legall,
Appl. Phys. Lett. {\bf 78}, 3663 (2001).

\bibitem{darwish04}
M. AlHajDarwish, {\it et al.,}
%H. Kurt, S. Urazhdin, A. Fert, R. Loloee, W. P. Pratt, Jr., and J. Bass,
Phys. Rev. Lett., {\bf 93}, 157203 (2004).

\bibitem{stiles02}
M. D. Stiles and A. Zangwill, Phys. Rev. B {\bf 66}, 014407 (2002);
J. Appl. Phys. {\bf 91}, 6812 (2002).

\bibitem{kiselev03}
S. I. Kiselev, {\it et al.},
%J. C. Sankey, I. N. Krivorotov, N. C. Emley, R. J. Schoelkopf, R. A. Buhrman and D. C. Ralph,
Nature {\bf 425}, 380 (2003).

\bibitem{krivorotov05}
I. N. Krivorotov, {\it et al.,}
%N. C. Emley, J.C. Sankey, S.I. Kiselev,  D. C. Ralph, and R. A. Buhrman,
Science {\bf 307}, 228 (2005)

\bibitem{xiao05}
J. Xiao, A. Zangwill, and M. D. Stiles, Phys. Rev. B {\bf 72}, 14446 (2005).

\bibitem{dietl02}
T.~Dietl, Semicond. Sci. Technol. {\bf 17}, 377 (2002).

\bibitem{ferrand01}
D.~Ferrand, {\it et al.,}
%A.~Wasiela, S.~Tatarenko, J.~Cibert, G.~Richter,
%P.~Grabs, G.~Schmidt, L. W.~Molenkamp, and T.~Dietl,
Sol. State Communs. {\bf 119}, 237 (2001).

\bibitem{rashba00}
E. I.~Rashba, Phys. Rev. B {\bf 62}, R16267 (2000).

\bibitem{gorley05}
P. M.~Gorley, {\it et al.,}
%P. P.~Horley, V. K.~Dugaev, J.~Barna\'s, and W.~Dobrowolski,
cond-mat/0508280 (2005).

\bibitem{sun00}
J. Z.~Sun, Phys. Rev. B {\bf 62}, 570 (2000).

\bibitem{brataas01}
A. Brataas, Yu. V. Nazarov, G. E. W. Bauer, Eur. Phys. J. B {\bf
22}, 99, (2001).

\bibitem{barnas05}
J.~Barna\'s, {\it et al.,}
%A.~Fert, M.~Gmitra, I.~Weymann, V.K.~Dugaev,
Phys. Rev. B {\bf 72}, 024426 (2005).

\bibitem{erugin74}
N. P.~Erugin, I. Z.~Shtokalo et al. {\em Ordinary Differential
Equations} (Vyshcha Shkola, Kiev, 1974).

\bibitem{haken83}
H.~Haken, {\em Advanced Synergetics} (Springer, Berlin, 1983).

\bibitem{gregg02}
J. F.~Gregg, {\it et al.,}
%I.~Petej, E.~Jouguelet and C.~Dennis,
J. Phys. D {\bf 35}, R121 (2002).

\bibitem{schmidt00}
G.~Schmidt, {\it et al.,}
%D.~Ferrand, L.W.~Molenkamp, A.T.~Filip and B.J.~van~Wees,
Phys. Rev. B {\bf 62}, R4790 (2000).

\bibitem{albrecht03}
J. D.~Albrecht and D. L.~Smith, Phys. Rev. B {\bf 68}, 035340
(2003).

\end{thebibliography}
\end{document}